\documentstyle[12pt,procsla]{article}
  
  \catcode`\@=11
  \long\def\@makefntext#1{
  \protect\noindent \hbox to 3.2pt {\hskip-.9pt  
  $^{{\ninerm\@thefnmark}}$\hfil}#1\hfill}		
  
  \def\@makefnmark{\hbox to 0pt{$^{\@thefnmark}$\hss}}  
  	
  \def\ps@myheadings{\let\@mkboth\@gobbletwo
  \def\@oddhead{\hbox{}
  \rightmark\hfil\ninerm\thepage}   
  \def\@oddfoot{}\def\@evenhead{\ninerm\thepage\hfil
  \leftmark\hbox{}}\def\@evenfoot{}
  \def\sectionmark##1{}\def\subsectionmark##1{}}
  
\begin{document}
  
  \centerline{\normalsize\bf STUDIES OF NEUTRINO OSCILLATIONS AT
REACTORS}
  \baselineskip=16pt
  
  \centerline{\footnotesize FELIX BOEHM}
  \baselineskip=13pt
  \centerline{\footnotesize\it Physics Department, California Institute of Technology,
  }
  \baselineskip=12pt
  \centerline{\footnotesize\it Pasadena, CA 91125, USA}
  \centerline{\footnotesize E-mail: boehm@caltech.edu}
  \vspace*{0.3cm}
  
  \vspace*{0.9cm}
  \abstracts{Experiments with reactor neutrinos continue to
shed light on our understanding of neutrino oscillations.
We review some of the early decisive experiments. We then turn to 
  the recent long baseline oscillation experiments at Palo Verde and Chooz 
which are leading to the conclusion that the atmospheric neutrino 
anomaly if attributed to oscillations does not 
involve an appreciable mixing with the $\bar\nu_e$ . 
  The very long baseline KamLAND experiment  
  is now in the planning stages. Its goal is to help explore the 
large mixing angle solar solution. A review of the
$\bar\nu_e + d$ experiment at Bugey 
and an outline of the $\bar\nu_e$ magnetic moment studies
complete this chapter.}
   
  \normalsize\baselineskip=15pt
  \setcounter{footnote}{0}
  \renewcommand{\thefootnote}{\alph{footnote}}

 \section{Introduction}

Neutrinos from reactors have played an important and  decisive 
role in the early
history of neutrino oscillations\cite{BV92}. 
After considerable controversy in the early 1980s,
results from the reactors at ILL\cite{Kw81} in 1981,
at Goesgen\cite{Za86} in 1986, and at Bugey\cite{Ac95} in 1995
have found no evidence for neutrino oscillations involving
reactor $\bar\nu_e$.
More recently, the Chooz\cite{chooz99} and the Palo Verde\cite{PV99}
experiments have confirmed these findings with greater sensitivity.
The purpose of this Chapter is to highlight the
developments involving reactor neutrinos and to outline the current 
status and future studies. 

We begin with a brief reminder of the parameters that play a role
in neutrino oscillation physics\cite{BV92}.
Assuming, for simplicity, that there are only
two neutrino flavors, then the two parameters describing oscillations 
are the mixing amplitude
$sin^2 2\theta$ and the mass parameter 
$\Delta m^2 $. They are related to the probability of creating a 
weak interaction state with flavor
$l'$ from a state $l$ $(l \neq l')$ in an "appearance experiment" through the
expression,

\begin{equation}
P_{ll'} = sin^22\theta ~ sin^2  \frac{1.27 \cdot \Delta m^2 (\rm eV)^2
\cdot L (\rm m)}{E_{\nu}(\rm MeV)} ~,
\end{equation} 
$L$ being the distance between neutrino source and detector ("baseline")
and $E_\nu$ the neutrino energy.
The probability that a state $l$ disappears through oscillation
is given by $P_{ll} = 1 - P_{ll'}$. While reactor $\bar\nu_e$ may 
oscillate into $\bar\nu_{\mu}$ or $\bar\nu_{\tau}$, these
neutrinos cannot be observed via charged current reaction as the
energy of the $\bar\nu_e$ at a reactor is insufficient to create a 
$\mu$ or a $\tau$.
A reactor experiment thus explores only
the disappearance of the $\bar\nu_e$.

As the oscillatory function depends on the ratio $L/E_\nu$,
it can be seen that low energy reactor neutrinos are well suited
in exploring the region of small
$\Delta m^2$ at relatively modest baselines.
For example, to explore the parameter $\Delta m^2$ down to $10^{-3} eV^2$
a reactor experiment with $E_{\nu}$ around 5 MeV requires a baseline
of $L$ = 1 km, while an accelerator experiment 
with $E_{\nu}$ = 5 GeV would require $L$ = 1,000 km.

It follows from (1) that oscillations manifest themselves through
modifications of the energy spectrum of neutrinos arriving in the
detector as well as by a change in the total neutrino yield.
Both of these aspects can be explored in an experiment.

Reactor experiments with their sensitivity to
small $\Delta m^2$, have been directed toward exploring
the physics of the atmospheric neutrino ratio\cite{Fu94}, a topic
described in Chapter 5 of this book.
If extended to even larger baselines, these experiments are 
capable of shedding 
light on the large mixing angle solar neutrino solution\cite{So98},
as discussed in Chapter 4.

Most reactor neutrino detectors are based on the 
interaction with the proton,

\begin{equation}
 \bar\nu_e + p = e^+ + n 
\end{equation}
with a threshold of 1.8 MeV. This inverse neutron decay has the
largest cross section among neutrino-nuclear reactions.
The presence of the time-correlated
$e^+ , n$ signature
provides a powerful way to retrieve the neutrino signal from the
abundant neutron and low energy radioactive backgrounds.
A small anisotropy of the reaction products arising from the 
kinematics of the detection process 
can be used for "pointing" and thus for background suppression.

  \section{The Reactor Neutrino Spectrum}

The neutrino sources for these experiments are large commercial
power reactors, each producing about 3GW of thermal power 
accompanied by
neutrino emission at a rate of about $8 \times 10^{20} \bar\nu_e /s$.
As a rule, these reactors run uninterruptedly at full power,
except for a refueling cycle of about 1 month per year
which provide opportunities for 
studying the backgrounds of the detector system.

The $\bar\nu_e$  spectrum from a fission reactor 
and its relation to the reactor's power and status in the burn cycle
is well understood today.
Pioneering work in deriving this spectrum taking into account
the contributions of the fissioning
isotopes $^{235}$U, $^{239}$Pu, $^{238}$U, $^{241}$Pu, and $^{252}$Cf
and their evolution during the burn cycle
has been reported by Vogel\cite{Vo81} in 1981. This extensive modeling work
has been
supplemented by experimental studies of the electron spectra of
the fissioning isotopes $^{235}$U, $^{239}$Pu, and $^{241}$Pu
with an on-line beta spectrometer at 
ILL Grenoble by Schreckenbach and others\cite{Sch89}.
The combined uncertainty for the predicted reactor neutrino spectrum
is about 3\%.

Figure 1 shows the time evolution of the reactor power associated
with the various fissioning fuel components, taken from ref\cite{Za86}.
This information was folded into the calculated neutrino
spectrum \cite{Vo81}.

\vspace*{1cm}
\fcaption
{Evolution of the contributions to the neutrino
spectrum by various reactor fuel components (from ref\cite{Za86}).}
\vspace*{0.6cm}

High statistics neutrino experiments involving the total neutrino yields
were carried out at the Bugey reactor\cite{De94}
at short distance from the reactor (where possible oscillation
effects are negligible).
A 2,000 $l$ water target was installed at a distance of 15 m from one of the
2,800 MW reactors at the Bugey site. As the detector responded to
neutrons only it provided an integral cross section of the reaction
$\nu_e + p = e^+ + n$  for neutrinos with energies
above the reaction threshold of 1.8 MeV.
The event rate was 3,021/d with a background rate (reactor off) of 2,600/d.
The cross section, obtained with an absolute accuracy of 1.4\%,
was compared to the calculated cross section based on V-A theory.
Moreover, these results confirm that the reactor neutrino spectrum
and its relation to reactor power and fuel composition is well understood.
The results are listed in Table 1 together with previous results
from Goesgen\cite{Za86} and Krasnoyarsk\cite{Ku91}.
There
is an excellent agreement between the measured and calculated neutrino rates
for all these experiments.

\begin{table*}[hbt]
\setlength{\tabcolsep}{1.5pc}
\newlength{\digitwidth} \settowidth{\digitwidth}{\rm 0}
\catcode`?=\active \def?{\kern\digitwidth}
\caption{Integral Cross Sections for Reactor Neutrinos on Protons}
\label{tab:effluents}
\begin{tabular*}{\textwidth}{@{}l@{\extracolsep{\fill}}cccc}
\hline
 & Goesgen(86) \cite{Za86} & Krasnoyarsk(90) \cite{Ku91} & Bugey(94) \cite{De94}\\ 
\hline
$\pm\sigma_{exp}$ & 3\% & 2.8\% & 1.4\% \\ 
$\sigma_{exp}/\sigma_{V-A}$ & 0.992 $\pm$ 0.04 & 0.985 $\pm$ 0.04 & 0.987 $\pm$ 0.03 \\ \hline 

\end{tabular*}
\end{table*}

The parameterization 
by Vogel and Engel\cite{VE89} serves as a convenient starting point
for present analyses of the measured reactor spectra.
As an example, Figure 2 shows the neutrino spectrum form
$^{235}$U fission together with the neutrino-proton reaction cross 
section and reaction yield.

\vspace*{1cm}
\fcaption
{Energy spectrum, cross section and yield of neutrinos from 
$^{235}$U fission in a reactor\cite{vogel}.}
\vspace*{0.6cm}

  \section{Oscillation Experiments}

\subsection{The ILL-Grenoble and Goesgen Experiments}

Motivated by theoretical developments of neutrino mass and
mixing in the 1970s\cite{Bi76} an early reactor experiment was
installed at the research reactor of the Laue Langevin Institute (ILL)
in 1977 with the aim of shedding light on oscillations involving $\bar\nu_e$. 
The neutrino detector in this ILL experiment\cite{Kw81}
consisted of 30 individual cells with liquid
scintillator which track the positron, sandwiched between  
$^3$He proportional chambers to detect the neutron. 
The distance between reactor and detector was 8.7m.
This "disappearance experiment" was searching for a possible
reduction of the $\bar\nu_e $ flux as well as for a modification of the
energy spectrum observed in the detector. It was found that
the measured neutrino spectrum agreed with that
calculated\cite{Vo81} and thus revealed no evidence for
oscillations down to $\Delta m^2$ = 0.15 $\rm eV^2$ for
$sin^2 2\theta \geq$ 0.25.

To enhance the sensitivity of this experiment
and to gain information on oscillations with smaller $\Delta m^2$, 
the ILL detector was modified and transferred to the more
powerful Goesgen reactor in Switzerland. 
Three experiments were carried out
between 1981 and 1985 with the detector at distances of 37.8 m, 45.9 m,
and 64.7 m from the reactor core.

The Goesgen detector\cite{Za86} consisted of an array of liquid scintillation
counters and $^3 \rm He$ multi-wire proportional chambers, surrounded
by an active scintillation veto counter and various shielding, as
illustrated in Figure 3. 

\vspace*{1cm}
\fcaption
{The Goesgen neutrino detector (from ref\cite{Za86})} 
\vspace*{0.6cm}

The signature of an event is given by a positron pulse in the liquid
scintillator followed by a neutron induced reaction in the $^3 \rm He $
counter. The time correlation and time window chosen are shown in Fig. 4.

\vspace*{1cm}
\fcaption
{Left: Distribution of time intervals between positron
and neutron in Goesgen detector. Right: Pulse shape spectra for reactor-on
(solid curve) and reactor-off (dotted curve). The peak to the left
represents the neutrino signal (positrons). The right peak is
caused by high energy cosmic ray induced neutrons (from ref\cite{Za86}).}
\vspace*{0.6cm}

Pulse shape discrimination was instrumental in reducing background
events associated with fast neutrons from cosmic rays, as illustrated
in Figure 4. No reactor associated backgrounds were seen, as could be
verified by comparing backgrounds with reactor-on and reactor-off.
The observed correlated positron spectra, corrected for detector
response and background, as a function of energy and position, 
are shown in Figure 5.
 
\vspace*{1cm}
\fcaption
{Positron spectra at three positions of the detector (from ref\cite{Za86}).
The solid curves are the predicted positron spectra for no oscillations
derived from fitting the reactor neutrino spectra to the data, while
the dashed curves are obtained from the calculated neutrino spectra.}
\vspace*{0.6cm}

In order to compare the spectra taken at various positions and at
different times, 
the relative reactor spectrum for each experiment had to be
known. Small differences
in reactor fuel composition were taken into account, although these
differences were minimized by conducting each experiment over a full
fuel cycle. Corrections for the relative difference in fuel
composition
varied by less than 5\%, with a negligibly small uncertainty. 

In the data analysis, the experimental positron spectra were compared to 
calculated spectra in two different ways.
First, an analysis (Analysis A) independent of the source neutrino 
spectrum was conducted. The neutrino spectrum was parameterized 
and a $\chi^2$, calculated for the difference between the experimental 
yield and the expected yield, was minimized 
for a fixed set of parameters $\Delta m^2 $ and $sin^2 2 \theta$.
A maximum likelihood test was used to obtain the 
exclusion plot shown in Figure 6.
In a second analysis (Analysis B) the measured spectra were compared to 
those calculated\cite{Vo81,Sch89} (also shown in Figure 4) with the 
results also shown in Figure 6.

\vspace*{1cm}
\fcaption
{Regions in the parameter space excluded at 90\% CL from the Goesgen
experiment\cite{Za86}. Analysis A is based on the ratios of the neutrino
spectra at three distances. Analysis B is based on the calculated
neutrino spectrum.}
\vspace*{0.6cm}

It is important to appreciate the sources of the limitations in 
the accuracy of the mixing angle sensitivity arising from the uncertainties
in the absolute normalization of the neutrino spectrum (3\%),  the
detector efficiency calibration using a calibrated neutron source
(4\%), the reaction cross sections (2\%), and the reactor power
(2\%), compounding to an uncertainty of about 6\%.

\subsection{The Bugey experiments}

A high statistics search
for neutrino oscillations at the 2,800 MW Bugey reactor with the detectors
at 15, 40, and 95m,
has been reported by the Bugey group\cite{Ac95} in 1995.
In this experiment, three identical
600 $l$ segmented detectors were used.
Each detector consisted of 98 prismatic cells
viewed by a 3-inch PM on each side. The cells were 
filled with $^6$Li loaded (0.15\%) scintillator. 
Two principal advantages were quoted for the $^6$Li loading:
the neutron capture time in the
scintillator is reduced to 30 $\mu$s (from about 170$\mu$s on protons), 
and the resulting $^3$H and $\alpha$-particles
can be distinguished from the reaction positrons by pulse shape
discrimination. The relative and absolute normalization errors
in these experiments were 2\% and 5\%, respectively.
The 95 m experiment provided the most stringent results for
the mass parameter $\Delta m^2$ of $10^{-2} eV^2 $. 

\subsection{The experiments at Rovno and Krasnoyarsk}

We mention briefly the experimental work at the Russian reactors at 
Rovno\cite{Af88} and Krasnoyarsk\cite{Vi94}.
At the Rovno power reactor a measurement of the total
neutrino induced neutron yield was carried out. 
The detector, at 18m from the reactor, consisted of a 
water target into which a large number
of $^3$He proportional counters were embedded. 
The observed neutrino yield agreed with that calculated to within 
about 3\%.  A similar experiment was carried out
at the three-reactor station at Krasnoyarsk.
There, the $^3$He neutron counters were embedded in polyethylene 
and stationed at 57m from reactors 1 and 2, and 231m from reactor 3.
While only total yields were obtained, by comparing rates
from reactors at different distances, information on oscillations
could be derived.  These detectors, however,  could not
provide information on the neutrino spectral shape. 
Also, these detectors possessed much higher inherent backgrounds
than the detectors discussed above which specify the reaction
neutron by a  $e^+ n$ correlated signal. 

\subsection{The long-baseline experiments at Palo Verde and Chooz}

  Results from atmospheric neutrino experiments, such as
  those from Kamiokande\cite{Fu94} have triggered reactor neutrino studies
  aimed at exploring the parameter region $\Delta m^2$ 
between $10^{-2}$ and
  $10^{-3}$ $eV^2$. Two experiments, both at $L$ around 1 km, 
have been conducted recently,
one at the French reactor station (2 reactors) at Chooz\cite{chooz99} 
and the other at the Palo Verde\cite{PV99} site in Arizona, USA (3 reactors).
Both experiments now have results and we 
describe them below. 
While these experiments were pursuing their goals,
the new Super-Kamiokande\cite{Fukuda2} results 
which appeared in 1998 
favor the $\nu_{\mu}$ - $\nu_{\tau}$ channel over 
$\nu_{\mu}$ - $\nu_e$ in some regions of the parameter plane. 

Another experiment in the Kamioka mine in Japan, referred to as
KamLAND\cite{kamland},
at a much larger distance from a number of power reactors, 
is still in the proposal stage. 

To illustrate the effect from oscillations on the positron spectrum,
Fig. 7 shows the expected spectrum for Chooz or Palo Verde
for the case of no oscillations as well as for the set of 
oscillation parameters favored by the Kamiokande results.
Clearly, the effects from Kamiokande-type oscillations on the 
spectrum should be quite pronounced.

\vspace*{1cm}
\fcaption
{Expected positron spectra for the Chooz or Palo Verde
experiments for "no oscillations" and for oscillations given by the
Kamiokande parameters.} 
\vspace{0.6cm}

The Chooz and Palo Verde 
experiments are based on the 
reaction $\bar{\nu_e} + p = e^{+} + n$ and rely on a 
($e^{+}$, $n$) correlated signature. Both experiments
make use of Gd loaded liquid scintillator.
Gd loading reduces the capture time owing to its large thermal neutron
capture cross section, and also gives rise to
a high energy gamma cascade of up to 8 MeV. Both features are valuable,
the short capture time helps reduce random coincidences and the large
gamma ray energy allows reduction of backgrounds as the energy threshold
can be set above that of radioactive decay products. In both experiments, the
amount of Gd dissolved in the scintillator is about 0.1\% by weight.
At a distance of ca 1 km from the reactor, the detector response is
about 5 events per day per ton of scintillator. The Chooz experiment takes
advantage of an existing deep tunnel reducing the cosmic ray muon background
substantially. The Palo Verde experiment being in a shallow underground
laboratory has to cope with a considerably larger muon rate and thus
has to rely on powerful background rejection.
Because of this, the two detectors are designed quite differently. 
The Chooz detector consists of a homogeneous
central volume of Gd scintillator, while the Palo Verde detector
is made from finely segmented detector cells.

\subsubsection{The Palo Verde Experiment}

The Palo Verde experiment\cite{PV99} is situated
near the Palo Verde nuclear power plant in Arizona (3 reactors, 11 GW 
thermal power).
The detector is installed in an underground cave
with 32 mwe overhead at a distance of $L$ = 890 m from reactors 1 
and 3, and 750 m from reactor 2.
Each reactor is shut down for refueling for a period of ca 40 days every year
providing the opportunity for establishing the detector background.

\vspace*{1cm}
\fcaption
{Schematic view of the Palo Verde detector. One of the cells with PMTs, oil
buffers, calibration LEDs and optical fiber flashers is shown lengthwise
at the bottom.}
\vspace{0.6cm}

\vspace*{1cm}
\fcaption
{Illustration of the neutrino reaction in the 
matrix of Gd loaded scintillator.}
\vspace{0.6cm}

The detector, shown schematically in Fig. 8, has a fiducial volume of
12 tons.  Its liquid scintillator, whose composition is
60\% mineral oil, 36\% pseudocumene, 4\% alcohol, and 0.1\% Gd, was developed in
collaboration with Bicron\cite{piepke}. It has an effective
light attenuation length of 10 m for 440 nm light. 
The detector consists of 66 cells, each 9 m long,
of which 7.4 m are active and 0.8 m on each end serve as an oil buffer. 
There is a 5 inch, low radioactivity photomultiplier
attached to each end, allowing both, the
anode and the last dynode to be read out. A blue LED installed at 0.9 m from
each PM, as well as optical fibers, allow each individual cell to be monitored.
A passive water shield, 1 m thick, surrounds the block of active cells to help
shield against radioactivity as well as muon induced neutrons. An active veto
counter consisting of 32 12-m long MACRO cells is placed on all 4 long sides
while a removable end-veto counter protects the ends of the cell matrix.

A diagram of the detector response showing the $\bar{\nu}_e$ reaction
and the gamma rays from Gd capture is given in Fig. 9.   

A neutrino signal consists of a fast (30ns) $e^+ \gamma \gamma$
trigger within a block of 3 x 5 cells, with the first hit having
$E \geq 500 $keV, and the second hit $E \geq 30 $keV. This second hit
includes the Compton response from the 511 keV annihilation gammas.
This fast triple coincidence is followed by a slow (200 $\mu s$) signal
associated with the 8 MeV gamma cascade following neutron capture in Gd
within a 5 x 7 scintillator cell matrix.
  
Energy calibrations could be carried out with the help of small sources that
were introduced through a set of Teflon tubes installed alongside a group
of detector cells. 
The response from these sources at various
positions made it possible to monitor the attenuation length of
the scintillator which exhibited only a negligible decline over the period
measured. Fig. 10 shows the light yield along the scintillator cell.
The PMT linearity was obtained with the help of a fiber optics flasher
while single photo-electron peaks were monitored with a blue LED.

\vspace*{1cm}
\fcaption
{Light yield along scintillator cell. The attenuation length of
the Gd scintillator is 10m.}
\vspace{0.6cm}

Inasmuch as the experiment aims at extracting absolute $\bar\nu_e$
induced reaction rates, knowledge of 
the detection efficiency is essential.
The positron efficiency was established with the help of the positron
emitter $^{22}$Na. (A calibrated $^{68}$Ge
source\cite{piepke1} dissolved in a special cell will also be implemented.)
To obtain the neutron efficiency, a calibrated AmBe source was used
in a tagged mode, i.e.
in coincidence with the 4.4 MeV gamma from $^{12}$C*.
From these calibrations, combined with Monte Carlo simulations
an average (over the detector) efficiency was obtained. For the 1999 run 
this efficiency was found to be 0.112 yielding a neutrino event rate
in the detector of 225 $\pm$ 8 per day.

From the 1998/99 data, the observed rates for a 147 d run with 
full reactor power (three reactors on) and a 54.7 d run with reduced power 
(two reactors on) yielded the positron spectrum 
shown in Fig. 11.
The time structure of the correlated signal is depicted in Fig. 12.
The measured decay time of 35$\mu$s agrees well with that 
modeled with a Monte Carlo simulation.

To test the oscillation scenario, a $\chi$-squared analysis in the
($\Delta m ^2 $ - $sin^2 2\theta$) plane was performed, taking into account
the small variations in $\bar\nu_e$ flux from the 
burn-up dependent fission rate of the reactor. 
The 90\% CL acceptance region was defined according to a
procedure suggested by Feldman and Cousins\cite{feldman}.
The data agreed well with
the no-oscillation hypothesis.
In an independent analysis, which does not rely on the "on" minus "off"
scheme, the intrinsic symmetry of
the dominant neutron background with respect to the time sequence
of the $e^+$ and $n$ signals was implemented to cancel a major
part of the neutron induced background.
A small neutron background that remained after subtraction of the
signal with reversed time sequence was
obtained from a Monte Carlo simulation of muon spallation\cite{YF}. 
Experimental data points
corresponding to energies $\geq$ 10 MeV
(beyond the positron energy spectrum) served to normalize the
calculated neutron spectrum. 
This analysis which was based on subtraction of the neutron 
background showed no 
evidence for $\bar\nu_e$ - $\bar\nu_X$ oscillations, and thus agreed
with the the results of the more traditional "on" minus "off" analysis.
The region in the parameter
space excluded at 90\% CL is depicted by the curve
"Palo Verde" in Fig. 16.

\vspace*{1cm}
\fcaption
{Correlated positron spectrum derived from a 3-reactor run  
and a 2-reactor run observed and expected for no oscillations. }
\vspace{0.6cm}

\vspace*{1cm}
\fcaption
{Decay time of the fourfold coincidence giving the neutron capture
time in our Gd-scintillator.}
\vspace{0.6cm}

The segmentation of the Palo Verde detector makes it possible to study the 
$\bar\nu_e$ - $n$ angular correlation of reaction (2).
This, in turn, establishes an independent
background determination.
From kinematics we find that the neutron moves preferentially in the direction
of the incoming neutrino, with an angular distribution limited by
\begin{equation}
\cos (\theta_{\nu,n})_{max} = 
[ 2\Delta /E_{\nu} - (\Delta^2 - m^2 )/E_{\nu}^2 ]^{1/2}\,,
\end{equation}
where $\Delta = M_n - M_p$.

From the Monte Carlo simulation it was found that the neutron 
scattering preserves
the angular distribution, resulting in a shift of the mean coordinate of the
neutron capture center $\langle x \rangle$ = 1.7 cm\cite{vogelbeacom}. The angular
spread after scattering is very pronounced as can be seen in Fig. 13.
It should be noted that this effect was
first studied by Zacek\cite{zacek} in connection with the segmented Goesgen
detector where the forward/backward ratio was found to be as large as a factor of 2.

Preliminary results\cite{Bo99} give an asymmetry expressed as events in the half
plane away from the reactor (forward) minus events in the half plane toward
the reactor (backward) of 109 $\pm$ 44, in agreement with a Monte Carlo
simulation.

\vspace*{1cm}
\fcaption
{Angular distribution of scattered (moderated) neutrons with 
regards to the neutrino direction.}
\vspace{0.6cm}

\subsubsection{The Chooz Experiment}

An experiment with a similar aim, however with a 
substantially different detector was
carried out at Chooz by a French-Italian-Russian-US collaboration.
This experiment and its results\cite{chooz99}, 
are reviewed below.

The Chooz detector is comprised of three regions, a central region containing 5 tons of
Gd loaded liquid scintillator and surrounded by an acrylic vessel, a containment
region with 17 tons of ordinary liquid scintillator,
and an outer veto region with 90 tons of scintillator.
Fig. 14 shows schematically the arrangement of the Chooz detector.

\vspace{1cm}
\fcaption
{Schematic arrangement of the Chooz detector}.
\vspace{0.6cm}

The inner two regions are viewed by a set of photomultipliers.
An independent set of PM detects the light from the veto region.
While the positron response is obtained from a signal
in the inner region, the neutron response comprises signals from the inner
region as well as from the containment region, resulting in a well contained
and well resolved Gd capture sum peak at 8 MeV.
As mentioned earlier, the Chooz detector is installed in
a tunnel, thus reducing the correlated background to less than 10\%
of the signal. 

The data was obtained at various power levels
of the two Chooz reactors as these reactors were slowly brought
into service.
A total of 2991 neutrino events was accumulated
in 8209 live hours for reactor-on, and 287 events in 3420 live hours
for reactor-off. Normalized to the full power of the two reactors
(8.5 GW th) the event rate corresponds to 27.4 $\pm$ 0.7 neutrino
interactions per day where the error includes contributions from
the reaction cross section, the reactor power, the number of 
protons in the target, and detector efficiency. In comparison,
the background rate was 1.0 $\pm$ 0.1 per day.
The ratio of measured-to-expected neutrino signal is 1.01 $\pm$ 2.8\% (st)
$\pm$ 2.7\% (sys).
The total efficiency of the detector was found to be 69.8 $\pm$ 1.1 \%. 

The positron energy spectrum for reactor-on
and reactor-off is shown in Fig. 15, together with a plot of the ratio of
measured-to-calculated spectrum.

The neutron capture event characterized by an 8 MeV gamma peak was
localized to within $\sigma _x$ = 17.4 cm. The energy resolution 
for the 8 MeV peak
was $\sigma _ e$ = 0.5 MeV, or about 1 MeV fwhm. Calibrations for 
energy, neutron efficiency, and timing, respectively,
were carried out with sources of $^{60}$Co, $^{252}$Cf, and AmBe.
The lifetime for neutron capture in the Gd scintillator was found
to be 30.5 $\mu$s. The combined systematic error was 2.8\%.

Figure 16 depicts the Chooz excluded area. Clearly, the Kamiokande region
is excluded with a high confidence level, implying the absence of
$\nu_e \leftrightarrow \nu_{\mu}$ oscillations.
The  mixing angle limit for large $\Delta m^2$ from this analysis is
$sin^2 {2\theta} <  0.1$
at 90\% CL, again based on the widely accepted method by
Feldman and Cousins\cite{feldman}.
The 90\% limit for $\Delta m^2$ for maximum mixing from this experiment
is 0.7 x $10^{-3} eV^2$.

\vspace{1cm}
\fcaption
{Positron energy spectra from the Chooz experiment.}
\vspace{0.6cm}

The Chooz collaboration has also compared the spectrum from reactor 2 which
is at $L$ = 998 m to that of reactor 1 at $L$ = 1115 m.
The relative spectra from the two reactors at different distances provided
information on oscillations independent of the absolute yields
as described above in the context of the "analysis A" of the
Goesgen experiment. For the Chooz experiment, that analysis
leads to an exclusion plot consistent with, however less stringent than,
that of their analysis involving absolute neutrino yields.

The Chooz analysis also 
includes a discussion\cite{choozangle} of the neutron angular 
distribution as mentioned in the section above on Palo Verde.

The parameter space that could be excluded by each of the aforementioned
experiments is summarized in Figure 16. The curves are labeled by 
the experiment and date providing a historical account of the
impressive gain in sensitivity over time.

\vspace*{1cm}
\fcaption
{An overview showing the evolution in time of the 90\% excluded regions 
for the experiments reviewed
in this paper. The Kamiokande allowed regions\cite{Fu94} and the region
allowed by the recent Super-Kamiokande\cite{Fukuda2} results if analyzed 
in the $\Delta m^2$ $vs.$ $\Theta_{1,3}$ \cite{Ok99} are also shown 
by the hatched areas.}
\vspace{0.6cm}

\subsection{The KamLAND Experiment}

The KamLAND\cite{kamland} experiment 
will be the ultimate long baseline reactor experiment, destined to explore
$\bar{\nu}_e$ disappearance at very small $\Delta m^2$. It will be sensitive
to exploring 
the large mixing angle solar MSW solution. The experiment will also address 
the small mixing angle solar
MSW at low neutrino energy by observing the $\nu_e$ - electron scattering, 
as well as, by invoking seasonal variations,
the vacuum oscillation solution.

The neutrinos originate from 16 nuclear power plants (130 GW thermal
power) at distances between 80 and 800 km
from the KamLAND detector, with 90\% of the neutrino flux produced
at sites between 80 and 214 km. The detector will be a 1 kT 
liquid scintillator to be
installed in the former Kamiokande cavity at a depth of 1,000 mwe. 
The spherical 1 kT scintillator and a surrounding 2.5 m mineral oil shielding
are contained in an 18 m diameter stainless steel sphere that also
supports the 2000 17-inch and 20-inch photomultipliers 
providing a 30\% light coverage.
The detector light yield is projected to be 100 photoelectrons
per MeV. A water volume surrounding the sphere serves as a 
Cerenkov veto counter. A schematic view of the detector is given in
Figure 17.  
The expected event rate associated with all power plants (51 reactors)
is projected to be 750/y. The background rates due to radioactivity and
neutrons from muon spallation were obtained from simulations  and
are predicted to be about 37/y. Under these conditions, a contour plot
can be constructed for a 3-year exposure which covers the large
mixing angle solar MSW solution, as shown in Figure 16,
with a maximum
sensitivity to $\Delta m^2$ of 4 x $10^{-6} eV^2$.
The KamLAND detector should be operational in 2001. 

\vspace*{1cm}
\fcaption
{Schematic view of the KamLAND detector (from ref\cite{kamland})}.
\vspace{0.6cm}

\section{The $\bar\nu_e - d$ experiment at Bugey}

Reactor neutrinos interacting with deuterons result in two reaction
channels, a charged current (CC) reaction

\begin{equation}
\bar\nu_e + d \rightarrow e^+ + n + n ,
\end{equation}
and a neutral current (NC) reaction

\begin{equation}
\bar\nu_e + d \rightarrow \bar\nu_e + p + n .
\end{equation}

As the CC reaction is sensitive to oscillations while the NC reaction 
is not, a measurement of the ratio of the
reaction yields may serve as a test for oscillations.
This scheme was first suggested by Reines et al.\cite{Re80}
and implemented at the Savannah River reactor in 1980.
However, owing to incomplete
understanding of the neutron efficiency for single and double neutron
events, the 1980 results turned out to be unreliable.
The experiment was 
repeated recently by Riley et al.\cite{Ri99}
essentially using the same 1980 apparatus
installed in the Bugey reactor in France.
The detector consisted of a central cylindrical volume containing
276 kg of deuterium
into which ten $^3$He proportional counters were immersed
serving as neutron detectors. A liquid scintillator veto as well
as Pb and Cd shielding surrounded the detector.
In this experiment, only neutrons were counted. 
In addition to the reaction neutrons, however, there was
a sizable number of background neutrons created by cosmic ray muons 
that could not be tagged with the veto.

The event rate for one-neutron events (NC) was 37.7 $\pm$ 2.0 per day
with a background rate of 57.0 $\pm$ 1.5 per day.
The rate for the two-neutron events (CC) was 2.45 $\pm$ 0.48 with a
background of 3.26 $\pm$ 0.36 per day.

After correcting for the efficiencies for detecting one neutron and 
two neutrons
of 0.29 $\pm$ 0.01 and 0.084 $\pm$ 0.006, respectively,
as determined with the help of a $^{252}$Cf source  and model calculations,
the ratio of the reaction yields for CC and NC divided by that calculated
was found to be 0.96 $\pm$ 0.23 consistent with 1, thus in
agreement with the prediction. It appears that this result differs 
substantially from their 1980 work\cite{Re80} with the same detector.

On account of the relatively large
statistical errors and the close distance (18.5m) to the reactor,
the experiment's sensitivity to oscillations was only modest
allowing it to exclude an area in the parameter space with
large mixing angles.
The work also confirmed the calculated cross sections describing the 
neutrino-deuteron break-up in the CC and NC channels.

\section{Neutrino Magnetic Moment}

If neutrinos have mass, they may have a magnetic moment.
An experimental effort to look for a neutrino magnetic moment, therefore,
is of great interest. Some indications for a magnetic moment have
come from a suggested correlation of the signal in the $^{37}$Cl experiment
with solar activity, suggesting a value of $10^{-11} - 10^{-10} \mu_{Bohr}$.
In addition, considerations of a possible resonant spin flavor precession (RSFP),
and also of neutrino interactions in supernovae have been mentioned.

The neutrino magnetic moment contributes to the $\nu_e e $ scattering\cite{vogelengel}
as shown in Fig. 18.  This contribution is most pronounced at low electron-recoil
energy. At about 300 keV the magnetic moment scattering is roughly equal
to the weak scattering.

Previous results by Reines et al.\cite{reines} from scattering
reactor neutrinos on electrons in a 16 kg plastic scintillator have given
$\mu_{\nu} = 2 -  4 \times 10^{-10} \mu_B$. More recently, Gurevitch et al.\cite{gurevitch},
using a 103 kg $C_6 F_6$ target found $\mu_{\nu} <  2.4 \times 10^{-10} \mu_B$,
and Derbin et al.\cite{derbin} with a 75 kg Si target 
found $\mu_{\nu} < 1.8 \times 10^{-10} \mu_B$.
A recent analysis\cite{beacomvogel}
of the Super-Kamiokande data\cite{Fu99} results in a
limit of $\mu_{nu} < 1.6 \times 10^{-10} \mu_B$.

\vspace*{1cm}
\fcaption
{Contribution of the neutrino magnetic moment to the $\bar{\nu}_e e \rightarrow
\bar{\nu}_e e $
scattering, averaged over the reactor $\bar{\nu}_e$ spectrum. The purely weak
cross section is also shown. (From Vogel and Engel\cite{VE89})}
\vspace*{0.6cm}

An effort to obtain a value or stringent limit of $\mu_{\nu}$ is now underway by the
MUNU experiment, a Grenoble-Munster-Neuchatel-Padova-Zurich
collaboration\cite{amsler} which has built a 1000 liter $CF_4$ TPC  
at 5atm (18.5 kg),
surrounded by an anti-Compton shield. This detector is now installed at the
Bugey reactor in France. The expected event rate in the interval of 0.5 to
1 MeV recoil energy is 5.1 per day, at an expected background of 4.5 per day.
Implementing the angular correlation of the scattered electrons with respect
to the incoming reactor neutrinos is expected to enhance the signal-to-noise
significantly.  A schematic view of the MUNU TPC is shown in Fig. 19.

\vspace*{1cm}
\fcaption
{Layout of the MUNU detector for the measurement of the neutrino
magnetic moment.}
\vspace*{0.6cm}

\section{Conclusion}

Reactor neutrinos with their low energies are well suited to explore
small $\Delta m^2$ for the $\bar{\nu_e}$ disappearance channel.
From the results of the Chooz and Palo Verde experiments it can be concluded
that the atmospheric $\nu_{\mu}$ deficiency cannot be attributed to $\bar{\nu}_{\mu} 
\leftrightarrow \bar{\nu}_e$ oscillations. The Chooz experiment has ruled
out this channel with large confidence level, and the first data set from
Palo Verde excludes it at 90\% CL. To improve on the mixing angle sensitivity in these
experiments so as to shed light on a possible 3-flavor solution
will be a challenging task. 
The KamLAND experiment at a very large baseline is now on the drawing board.
Searches for a neutrino magnetic moment from the MUNU experiment are in progress.


\begin{thebibliography}{99}

\bibitem{BV92} See, for example, F. Boehm and P.Vogel, {\it Physics of Massive
Neutrinos}, Second Edition, Cambridge University Press (1992).

\bibitem{Kw81} H.Kwon et al., {\it Phys.Rev.D} {\bf 24} (1981) 1097.

\bibitem{Za86} G. Zacek et al., {\it Phys.Rev.D} {\bf 34} (1986) 2621.

\bibitem{Ac95} B. Achkar et al., {\it Nucl.Phys.B} {\bf 434} (1995) 503.

\bibitem{chooz99} M. Apollonio et al., {\it Phys.Lett.B} {\bf420} (1998) 397;
M. Apollonio et al., {\it Phys.Lett.B} {\bf 466} (1999) 415.


\bibitem{PV99}  F. Boehm et al., {\it Phys.Rev.Lett.}, {\bf84} (2000)
3764, F. Boehm et al. {\it Phys.Rev.D}, to be published; hep-ex/0003022.

\bibitem{Fu94} Y. Fukuda et al., {\it Phys.Lett.B}  {\bf
335} (1994) 237.

\bibitem{So98}  J.N.~Bahcall, P.I.~Krastev and A.Y.~Smirnov, 
{\it Phys.Rev.D} {\bf 58}, 096016 (1998).



\bibitem{Vo81} P. Vogel et al., {\it Phys.Rev.C} {\bf 24} (1981) 1543.

\bibitem{Sch89} K. Schreckenbach et al., {\it Phys.Lett.B} {\bf 218} (1989) 365.

\bibitem{De94} Y. Declais et al., {\it Phys.Lett.B} {\bf 338} (1994) 383.

\bibitem{VE89} P. Vogel and J. Engel, {\it Phys.Rev.D} {\bf 39} (1989) 3378.

\bibitem{vogel} P. Vogel, private communication.

\bibitem{Bi76} See, for example, S. M. Bilenky and P. Pontecorvo,
 {\it Physics Report} {\bf 41} (1978) 225, H. Fritzsch and P. Minkowski,
{\it Phys. Lett. B} {\bf 62} (1976) 72.

\bibitem{Ku91} A. A. Kuvshinnikov et al.,{\it JETP Lett.} {\bf 54} (1991) 255.

\bibitem{Af88} A. I. Alfonin et al., {\it JETP} {\bf 67} (1998) 213.

\bibitem{Vi94} G. S. Vidyakin et al., {\it JETP Lett.} {\bf 59} (1994) 390.

\bibitem{Fukuda2}  Y. Fukuda et al., {\it Phys.Rev.Lett.} {\bf 81}
(1998) 1562.

\bibitem{kamland}  P. Alvisatos et al., {KamLAND, a Liquid Scintillator
Anti-Neutrino Detector at Kamioka}, Stanford-HEP-98-03, Tohoku-RCNS-98-15
July 1998 (unpublished); A. Piepke, in {\it Neutrino 2000, International
Conference on Neutrino Physics and Astrophysics}, Sudbury, Canada,
June 16 - 21, 2000.

\bibitem{piepke} A. Piepke at al., {\it Nucl.Instr.Meth.A} {\bf 432}
(1999) 392.

\bibitem{piepke1} A. Piepke and B. Cook, {\it Nucl.Instr.Meth.A} {\bf 385}
(1996) 85.

\bibitem{feldman} G.J. Feldman and R.D. Cousins, {\it Phys.Rev.D} {\bf 57}
(1998) 3873.

\bibitem{YF} Y-F. Wang, L. Miller, and G. Gratta, {\it Phys.Rev.D}
{bf 61} (July 2000) to be published; hep-ex/0002050.
\bibitem{vogelbeacom} P. Vogel and J. F. Beacom, {\it Phys.Rev.D}
{\bf 60} (1999) 053003; K.B. Lee, private communication.

\bibitem{zacek} G. Zacek, Thesis, Technical University of Munich (1984).

\bibitem{Bo99} F. Boehm, in {\it Eighth International Workshop on
Neutrino Telescopes}, Venice, Feb 23 - 26, 1999, Ed. M. Baldo Ceolin, 
Edizioni Papergraf (1999).

\bibitem{Ok99} K.Okumara, PhD Thesis, University of Tokyo, unpublished,
and Super-Kamiokande Collaboration, preliminary results. See
also second quoted Ref. 6.

\bibitem{choozangle} M. Apollonio et al., {\it Phys.Rev.D} {\bf 61}
(2000) 012001.

\bibitem{Re80} F. Reines et al., {\it Phys.Rev.Lett.} {\bf 45} (1980) 1307.

\bibitem{Ri99} S. P. Riley et al., {\it Phys.Rev.C} {\bf 59} (1999) 1780.

\bibitem{vogelengel} P. Vogel and J. Engel, {\it Phys.Rev.D} {\bf 39} (1089)
3378.

\bibitem{reines} F. Reines et al., {\it Phys.Rev.Lett.} {\bf 37} (1976) 315.

\bibitem{gurevitch} I.I. Gurevitch et al., {\it JETP Lett.} {\bf 49} (1989) 740.

\bibitem{derbin} A.I. Derbin et al., {\it JETP Lett.} {\bf 57} (1993) 768.

\bibitem{Fu99} Y. Fukida et al., {\it Phys.Rev.Lett.} {\bf 82} (1999) 2430.

\bibitem{beacomvogel} J. F. Beacom and P. Vogel, {\it Phys.Rev.Lett.} {\bf 83}
(1999) 5222.

\bibitem{amsler} C. Amsler et al., {\it Nucl.Instr.Meth.A} {\bf 396} (1997) 115.

\end{thebibliography}
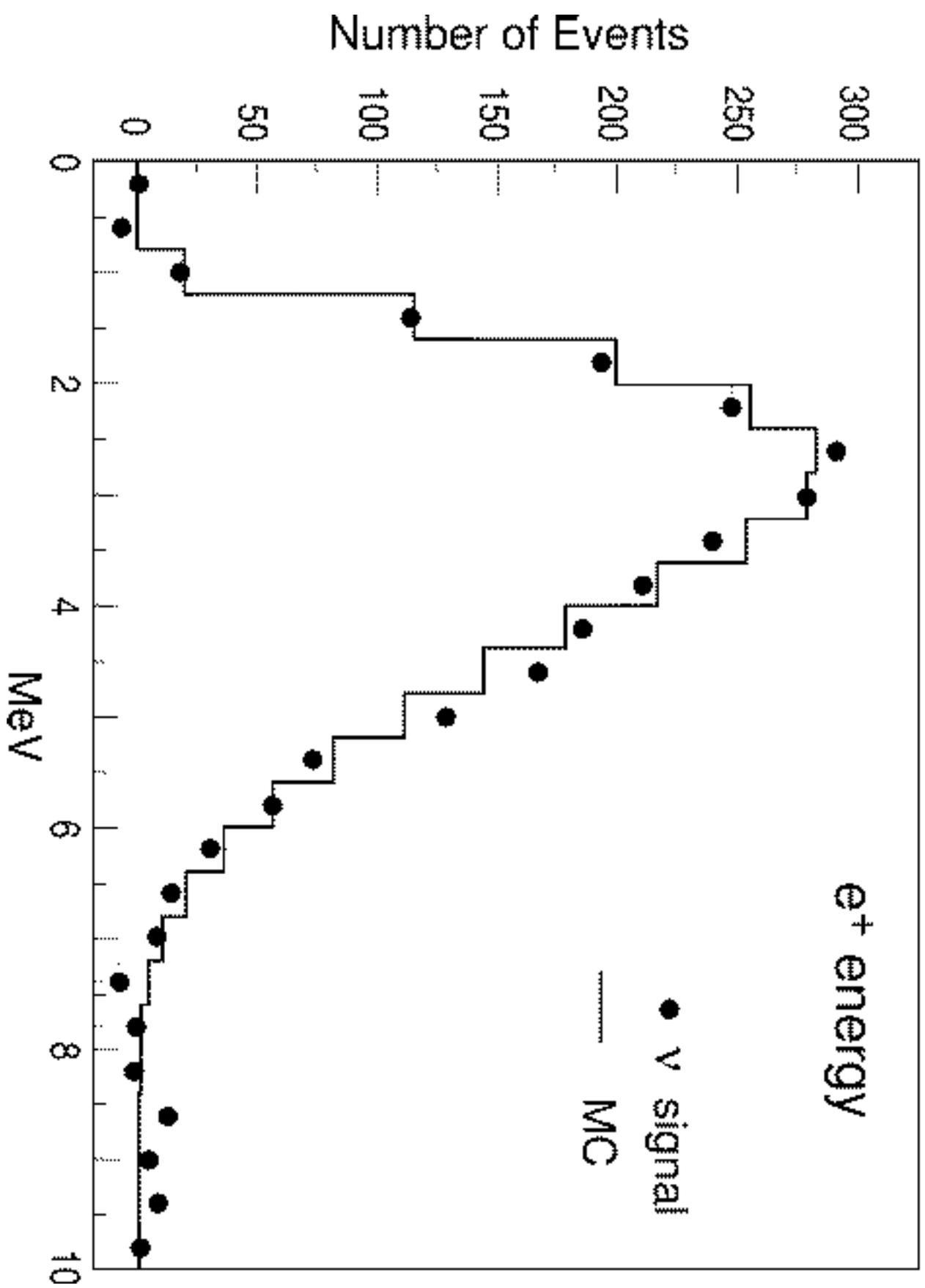
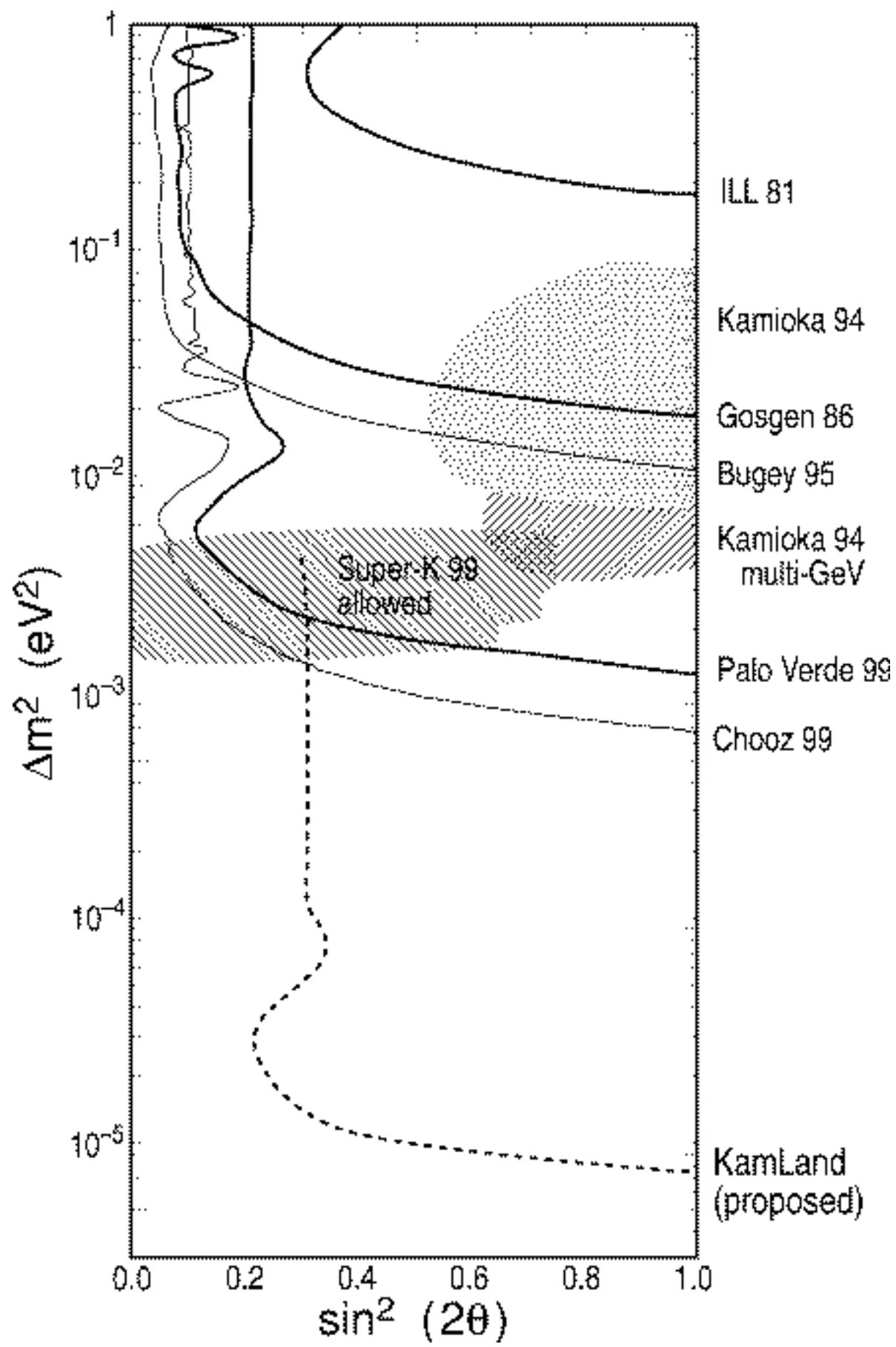
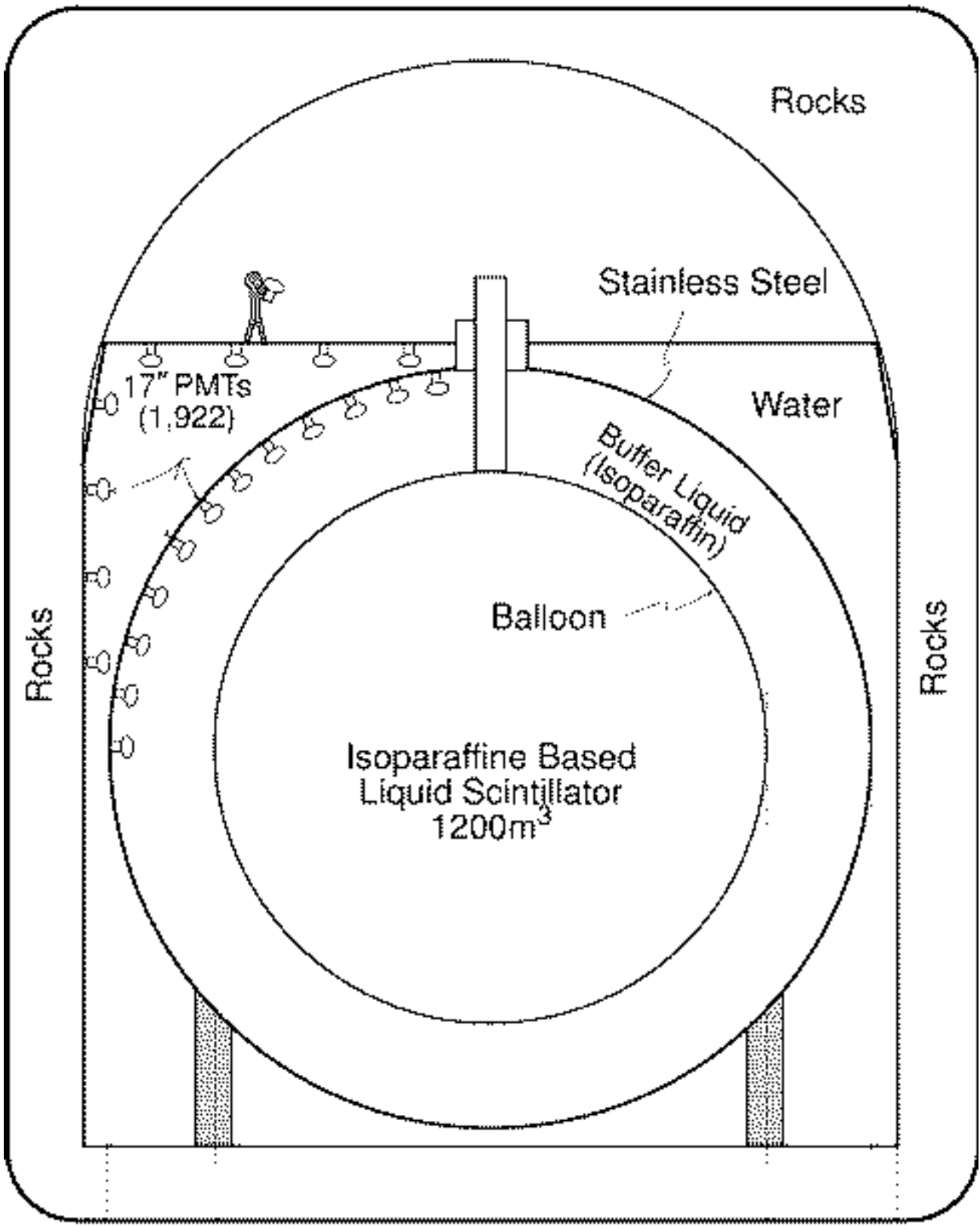
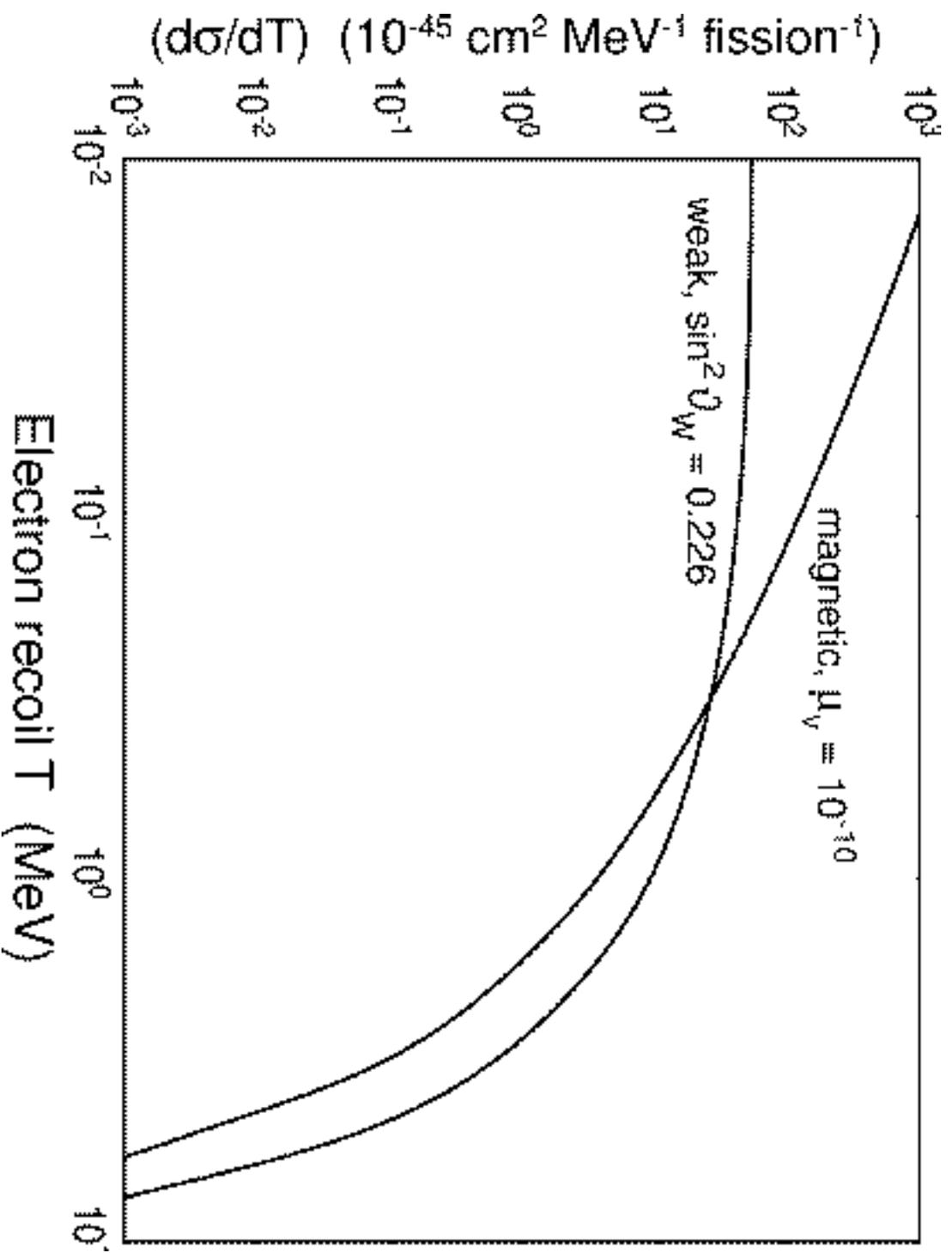
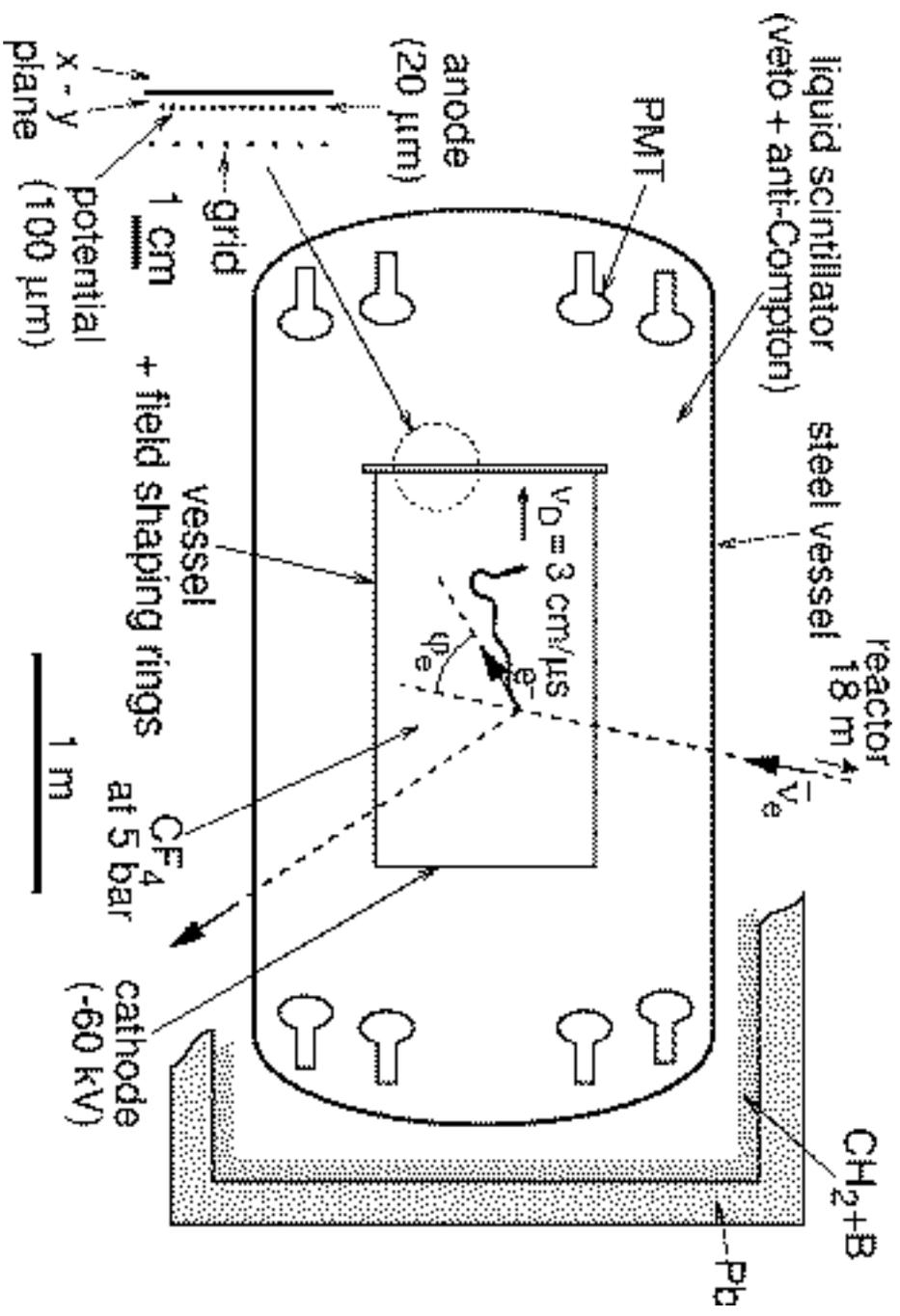
\end{document}